Regular Research Paper Accepted by FECS'21 (The 17th Int'l Conf on Frontiers in Education: Computer Science and Computer Engineering)

# The Study of Peer Assessment Impact on Group Learning Activities

Zhiyuan Chen*
School of Computer Science
University of Nottingham Malaysia
Jalan Broga, Semenyih, Selangor.
zhiyuan.chen@nottingham.edu.my

Soon Boon Lee
School of Computer Science
University of Nottingham Malaysia
Jalan Broga, Semenyih, Selangor.
lsboon99@gmail.com

Shazia Paras Shaikh
School of Computer Science
University of Nottingham Malaysia
Jalan Broga, Semenyih, Selangor.
shaziaparas20@gmail.com

Mirza Rayana Sanzana
School of Computer Science
University of Nottingham Malaysia
Jalan Broga, Semenyih, Selangor.
chrislam1112@gmail.com

**Abstract**

Comparing with lecturer marked assessments, peer assessment is a more comprehensive learning process and many of the associated problems have occurred. In this research work, we study the peer-assessment impact on group learning activities in order to provide a complete and systematic review, increase the practice and quality of the peer assessment process. Pilot studies were conducted and took the form of surveys, focus group interviews, and questionnaires. Preliminary surveys were conducted with 582 students and 276 responses were received, giving a response rate of 47.4%. The results show 37% student will choose individual work over group work if given the choice. In the case study, 82.1% of the total of 28 students have enjoyed working in a group using Facebook as communication tools. 89.3% of the students can demonstrate their skills through group-working and most importantly, 82.1% of them agree that peer assessment is an impartial method of assessment with the help of Facebook as proof of self-contribution. Our suggestions to make group work a pleasant experience are by identifying and taking action against the freeloader, giving credit to the deserving students, educating students on how to give constructive feedback and making the assessment process transparent to all.

**Keywords:** peer assessment, group learning, survey, focus group, case study

# 1. INTRODUCTION

In the University of Nottingham Malaysia, school of computer science, peer-assessment has been used with many modules (such as G52HCI, G54IHC, and G52GRP) since 2012. It involves students in the learning process and allows students to make judgments about each other's achievements. However, comparing with lecturer marked assessments, peer assessment is a more comprehensive learning process in some ways and many of the associated problems have occurred. Therefore, we conduct a research work to study the peer-assessment impact on group learning activities. The main aims of this study are 1) to provide a complete and systematic review of this field; 2) to increase the practice and quality of the peer assessment process; 3) and finally to improve the effectiveness of students learning in groups with a proper peer assessment system. The success of peer-assessment schemes depends greatly on how the process is set-up and subsequently managed. In this research, pilot studies were conducted and took the form of surveys, focus group interviews, and questionnaires that examined the impact of using peer assessment to improve students learning within groups.

Peer-assessment is seldom used by some lecturers due to concerns about the validity and reliability, problems of inaccuracy/low precision of student markers. However, there is considerable evidence that students can peer assess effectively (Nagori & Cooper, 2014; Nicol & Macfarlane-Dick, 2006). Several researchers have provided guidelines for best practices for the management of peer-assessment (Nagori & Cooper, 2014; Nicol & Macfarlane-Dick, 2006; Poon, 2011; Samantha, 2012). This study has experimented with the use of these best practice schemes. According to (Nicol & Macfarlane-Dick, 2006), the model of self-regulated learning and the feedback principles to develop self-regulation in students recommends that learners are actively involved in monitoring and regulating their performance through internal and external feedbacks. A Focus group can be conducted either in a traditional manner or online (Brüggen & Williams, 2018). This study considered several dimensions and criteria suggested in Assessment Standards knowledge exchange (ASKe), a center for excellence in teaching and learning based in the business school at Oxford Brookes University.

# 2. RELATED WORK

Poon (2011) designed a peer assessment approach to evaluate individual contribution in group work of the project module. Surveys and interviews were considered as data gathering tools. The results concluded that accord-ing to the participants' view, the peer evaluation method was fair and it maximizes the value of the group project. Gatfield (2006) conducted a detailed study to evaluate student satisfaction with the assessment process. The results showed that students were satisfied with the group work and the assessment method. It was also found that gender and age factors did not affect the satisfaction level. A substantial positive difference was found in the perceptions of international students (N = 90) in comparison with Australian students (N = 171) and in those students who did not have prior work experience. According to (Burnett & Brailsford), it is important to note that peer-assessment should not be relied on as a sole means of grading individuals, it should be used in conjunction with other assessment methods based purely on a student’s specific work, e.g. essays/reports, exams.Researchers also found out that students obtained behavioral, intellectual and personal skills from group work experience (Yen, et al., 2009; Hassanien, 2006). Hassanien (2006) conducted a study to explore the feelings and experiences of students regarding group work and group assessment. The study showed that group work provided invaluable experience that developed students' learning and achievement although several students disagree that peer assessment as a tool to enhance learning.

According to Topping (2009), generally peer assessment helped equal-status learners evaluate each other’s performance.
The main motive of peer assessment mentioned by several researchers is:

- To help students manage their learning abilities by getting feedback from their peers (Liu & Carless, 2006).
- To improve students learning as well as their group work (Topping, 2009).
- Peer assessment motive is not to enhance the evaluation or scores of the students but the learning process (Tillema, et al., 2011).

Focus group interview is a data collection method (McLafferty, 2004), it has two main components that make it successful, i.e. interaction and active role performed by the moderator (Barbour, 2007). Focus groups are not ordinary groups; focus group is meant for data collection while other ordinary groups are used for other purposes such as decision making or for academic purposes (Kamberelis & Dimitriadis, 2005; Ho, 2006). The benefit of using a focus group is that it provides more control and in-depth insight of the situation as it is audio or video recorded. (Krueger & Casey, 1994; Hofmeyer & Scott, 2007; Kitzinger, 1994; Warr, 2005) have stated that the main advantage of a focus group is that, in a normal social setting, it permits the people to interact with whom they already know. A Focus group can be conducted either in a traditional manner or online. Brüggen and Williams (2018) preferred a traditional focus group more as it provides more interaction and a detailed amount of data than online focus group interviews. Focus group interviews not only help participants to share their views and experiences but also provide a chance for the participants to hear and question other opinions as well (Hyde, et al., 2005). Through the focus group, participants have an opportunity to get clarification on a certain point, can ask for further explanation, and can get a clear insight. Through these discussions during the focus group, it is more likely to have healthy data and understanding (Kitzinger, 1994).

Yang and Tsai (2010) mentioned paper-based peer assessment has fewer advantages than online peer assessment. Application of technology makes the assessors comment freely without any restriction of time and location, also providing a chance of direct connection between them (Boud, et al., 2014). In the current 21st century, the use of technology is very rapid. On the Internet, social networking sites are most visited from all around the world specially by the young generation (Shih, 2011).

The rapid development of Internet technologies has ushered in an increasing interest in online web learning (Lin, et al., 2001; McCarthy, 2010). Facebook is one of the technologies which allows users to interact and collaborate, but it is not extensively used in tertiary education. Facebook, social networking site, allows students to engage in conversation, questioning, commenting, and passing opinions on each other's work, which are aligned with the social constructivist theory (Shih, 2011). According to Shih (2011), pedagogy, technology, and social interaction are the key factors for a technology-enhanced learning environment. Integrating social media (Facebook) with blended learning in higher education seems to be a feasible means for teachers to enhance students learning. Another reason for adopting the Facebook platform is, according to the statistics highlighted in Global social networks ranked by users (Clement, 2020). This statistic shows that Facebook is a commonly used social media website nowadays than any other social networking platform such as Twitter.

(Lai & Hwang, 2015) recognized that immersing students in higher-order thinking helped in boosting their skills and judgment and suggested a reciprocal peer assessment technique to aid the students to cultivate the marking criteria and educate themselves from examining their peers' works. (Staubitz, et al., 2016) observed the peer assessment method closely as it was believed to have an impact on personal feedback and interactive coursework along with being expandable, but on the downside it has the issue of the accuracy in grading and favoritism. MOOC has adopted peer assessment (PA) to handle the issues of grading creativity and performance using an automated machine grading system.

Reinholz (2016) suggested a model of an assessment cycle which distinguished the difference between self-assessment and peer-assessment and went further to explore learning contingencies. Mogessie (2017) highlighted that despite studying peer assessment for years, it did not make a significant proposition because of the challenges that came up from the manual process of peer assessment factor. Gao *et al.* (2019) suggested to reward the authors and the reviewers for a good understanding of problems and a proposal of a good solution that would portray the advantage of peer assessment and transparency of it.

Patchan *et al.* (2017) investigated the depth of peer assessment by having a comparison of the three forms of conditions involving accountability which were rating accountability, feedback accountability, and the combination of rating and feedback. Monte Carlo Simulation method has been implemented to show the effective output of peer-assessed work using the ascertained grades of peer assessment (Babik, et al., 2019). Online peer assessment has determined adequate feedback within a short period as compared to the classical methods, and hence several online peer review and assessment (OPRA) have been escalated to enable the automation of peer assessment and augment the procedure (Babik, et al., 2016; Babik, et al., 2017; Song, et al., 2017).

Pandero (2016) has validated two definite forms of using PA to improve the knowledge and learning process where the former one involving the assesse of the student whose work was evaluated by other students or peers and that individual obtains the criticism and comments from his peers and the latter involving the assessor who gained more awareness about his and his peers' strengths and weaknesses while evaluating the work of his peer. Adachi *et al.* (2017) highlighted the advantages of peer assessment through an interview process of academics which strongly portrayed the self and PA as a powerful tool in constructing students mind positively. Flachikov (2001) showed evidence that peer feedback enhanced student learning and proved the effectiveness by integrating peer assessment with technology i.e. via social media, Facebook.

## 3. METHODOLOGY

In this research, pilot studies were conducted and took the form of surveys, focus group interviews, and questionnaires that examined the impact of using peer assessment to improve students learning within groups. Preliminary surveys were conducted with 582 students from G52HCI, G51SYS, H83PDC, F40FMT, and MM1TF1 modules at the beginning of the class where 276 responses were received (Figure 1). To further understand issues identified in the survey and collect students' viewpoints, the focus group interviews were conducted in the school of computer science with 12 participants (all had peer assessment experience, 10 undergraduate students and 2 Ph.D. students). At the end, a case study for HCI coursework 2 with a peer-assessed group project using Facebook for team collaboration, feedback and monitoring has been conducted.

Before starting the session, the objectives of the session were formally explained to the students. The students were later asked to fill up the consent forms and were told that the entire session will be video recorded.

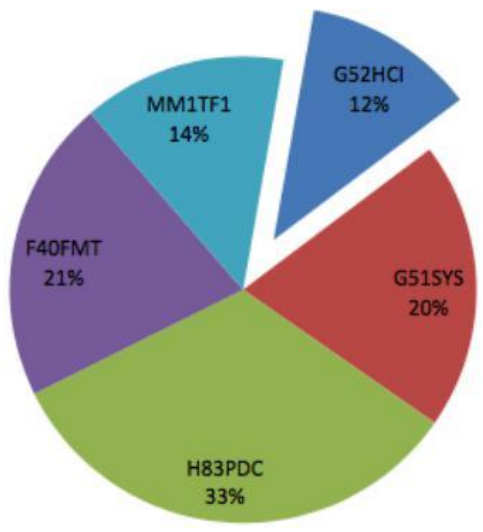

| | |
|---|---|
| G52HCI | 33 |
| G51SYS | 55 |
| H83PDC | 91 |
| F40FMT | 58 |
| MM1TF1 | 39 |
| total: | 276 |

Figure 1: Students distribution from the survey

The focus group session was divided into four phases: introduction, engagement, exploration, and an exit phase. During the introduction phase, the participants were briefly explained about how the session was going to run and how it would be recorded. The confidentiality issues and rules were also emphasized. Furthermore, the purpose, objectives, and outcome of the session were presented as well.

During the engagement phase, the participants were asked general questions initially, so that they become comfortable in answering questions. In the exploration phase, participants were asked detailed questions regarding the topic. In the exit phase, the participants were asked for additional comments.

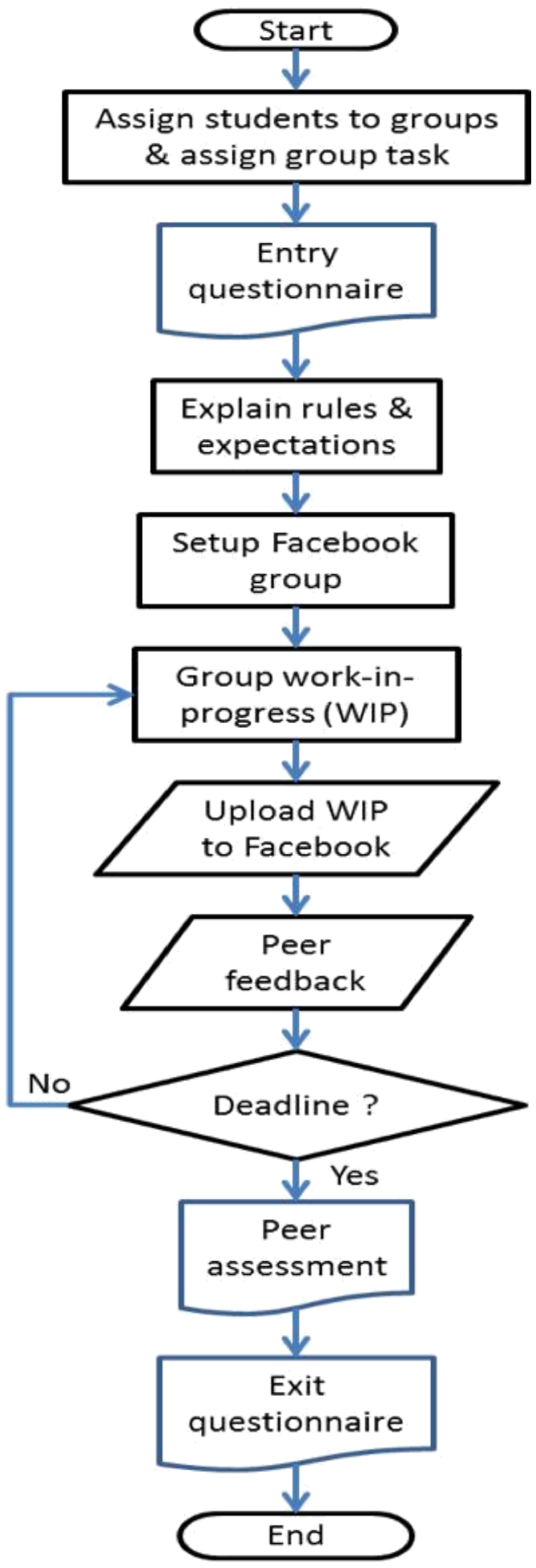

Figure 2: HCI Coursework 2 FlowChart

The case study of a peer-assessed group project for HCI Coursework 2 using Facebook for team collaboration has been conducted, where 27 questionnaires were returned and analyzed. Flowchart of the case study is shown in Figure 2. More than 80% of students stated they enjoyed working as a group and believed peer-assessment was a fair method of assessment. However, this case study is only with a small specific population, actual solutions could be more complicated, and more specific support could be needed.

## 4. RESULTS AND DISCUSSION

Preliminary surveys were conducted with 582 students from G52HCI, G51SYS, H83PDC, F40FMT, and MM1TF1 modules at the beginning of the class. Fourteen questions were asked, and 276 responses were received, giving a response rate of 47.4%. Figure 3 shows the overall results of the survey.

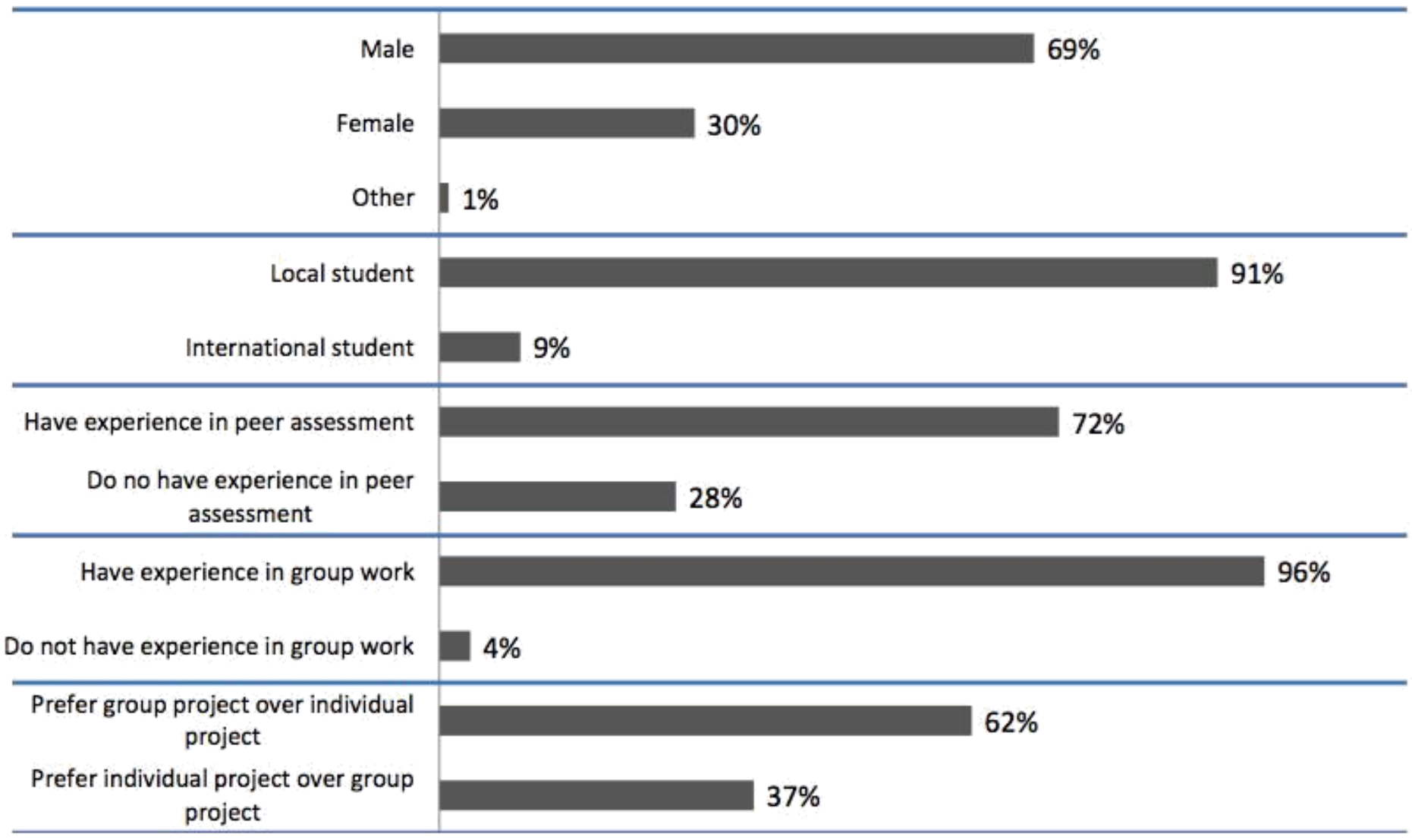

Figure 3: Survey Result 1

Given the choice, 37% will choose individual work over group work. The main reasons why student dislike group work are found to be in the conflicts within the group and members' unequal contribution. Details have been shown in Figure 4. Firstly, an individual's perception of fairness is an issue. 47% of the respondents said that they would be fair only if they perceive that others will also be fair, and only 34% have faith that others are generally fair. Most people justify their ill-will as a fairness issue. There is no doubt that unfairness exists and that some amount of it can never go away, especially because perfect fairness is impossible to define. Secondly attitude towards peer assessment also plays a significant role. According to the result, 63% of students are comfortable with assessing their peers, 18% feel that they are not qualified, whereas 18% do not feel comfortable with assessing their peers. Also,74% of students' value feedback from their peers, while 17% said that it is upsetting to receive negative feedback, and 8% said that it is irrelevant. When giving negative feedback, 49% of the students stated that they would be objective, and 50% will say as little as possible because they do not feel good about relaying negative feedback. Several group dynamics involve a group of students striving for common objectives where 56% of the student will assist other members after they have completed their work for the benefit of the group. On the other hand, 72% of students will consult other members for a solution known as disruptive behavior in the group. Also, 29% of students will place their desire for a group consensus above their desire to reach the right decision; 23% will argue persuasively for that which they feel is right; 17% will impose their views on the group, and 28% will seek help from outside of the group.

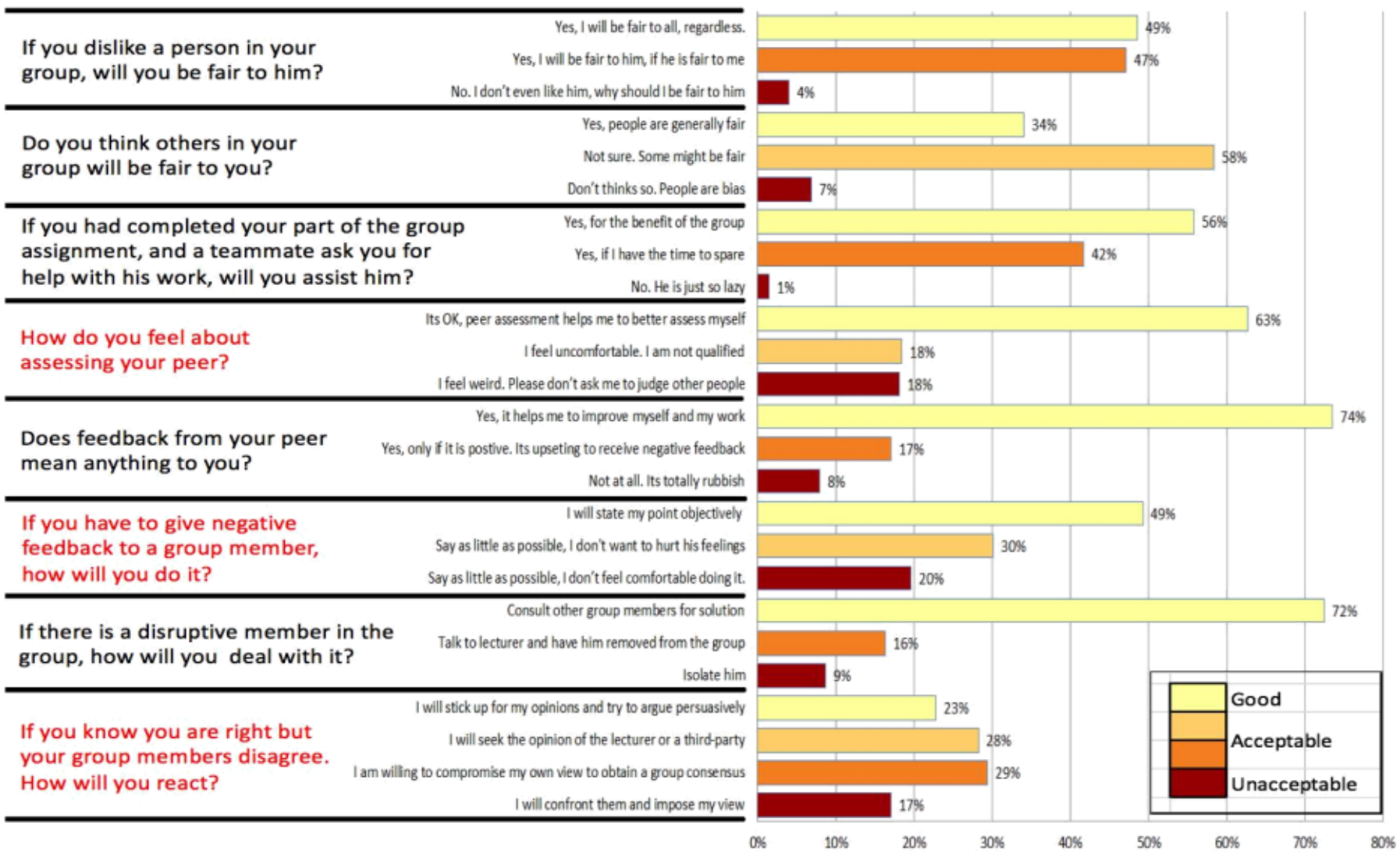

Figure 4: Students attitude and behavior towards group work & peer assessment

Disruptive behavior that can negatively affect the group dynamics are classified as (1) Aggressor, who often disagrees with others or is inappropriately outspoken; (2) Negator, who is often critical of others' ideas; (3) Withdrawer, who does not participate in the discussion; (4) Recognition seeker, who is boastful or dominates the session and (5) Joker, who introduces humor at inappropriate times.

Conflict is inevitable in groups and can be a very positive experience if managed appropriately. Group conflicts most likely occur in the brain-storming stage. Conflicts generate creativity in the problem-solving process and can lead to better results as people tend to sit back and say nothing when appropriate to avoid conflict. Compromising to obtain a group consensus is the middle ground between competing and accommodating. The greater good should always be the priority.

To address the above issues, we suggest making group work a pleasant experience by identifying and acting against the freeloader and giving more credit to the deserving students. Also, educate students on giving constructive feedback where feedback needs to be constructive instead of negative. Moreover, to make the assessment process transparent to all and at the same time, making it easier for tutors and peers to give a formative assessment.

All participants with mutual discussion mentioned that it is difficult to put a person into a criterion, peer assessment creates many conflicts between two persons. Furthermore, the participants mentioned that usually there is only one person working in a group so, all members attack that one person easily for the marks. The other group members get a higher score even though they work more than them, group leaders get targeted during the peer assessment process.

| | | | | |
|---|---|---|---|---|
| **A Teaching & Learning Questionnaire - group work & peer assessment** | | | | |
| **Exit Questionnaire Result** | | **TOTAL STUDENTS : 28** | | |
| | | | Total | % |
| 1 | Have you enjoyed working in a group? | Yes | 23 | 82.1 |
| | | No | 4 | 14.3 |
| 2 | Do you think you were able to demonstrate your skills through group-working? | Yes | 25 | 89.3 |
| | | No | 2 | 7.1 |
| 3 | Did working in a group motivate you ... | More | 9 | 32.1 |
| | | Less | 1 | 3.6 |
| | | the Same | 17 | 60.7 |
| 4 | Overall, how effectively did your team work together on the group assignment? | Extremely well | 4 | 14.3 |
| | | Well | 17 | 60.7 |
| | | Adequately | 6 | 21.4 |
| | | Poorly | 0 | 0.0 |
| 8 | Do you think peer assessment is a fair method of assessment? | Yes | 23 | 82.1 |
| | | No | 4 | 14.3 |
| 9 | Given the choice, would you participate in peer assessment again? | Yes | 22 | 78.6 |
| | | No | 5 | 17.9 |
| 10 | Do you think peer assessment helps you understand more about the assignment and what is required of | Yes | 8 | 28.6 |
| | | No | 6 | 21.4 |
| | | Not sure | 11 | 39.3 |
| 11 | Did you receive sufficient feedback from your group members on your contribution? | Yes, more than enough | 2 | 7.1 |
| | | Yes, sufficient | 14 | 50.0 |
| | | No, not much comment | 9 | 32.1 |
| | | No feedback at all | 2 | 7.1 |
| 13 | Did using Facebook make your group members' contribution more transparent? | Yes | 23 | 82.1 |
| | | No | 2 | 7.1 |
| | | Not sure | 2 | 7.1 |
| 14 | Did using Facebook help increase your interaction and participation in the group assignment? | Yes | 19 | 67.9 |
| | | No | 2 | 7.1 |
| | | the Same | 6 | 21.4 |

Figure 5: Exit Questionnaire Result

According to the exit questionnaire result in Figure 5, 82.1% of the total of 28 students have enjoyed working in a group using Facebook as communication tools. 89.3% of the students can demonstrate their skills through group-working and most importantly, 82.1% of them agree that peer assessment is an impartial method of assessment with the help of Facebook as proof of self-contribution. The main reason behind this is 82.1% of the students mentioned that by using Facebook, it provides transparency for each group member's contribution leading to 78.6% of the students will participate in the peer assessment again.

Participants suggested that it would be better if an online peer assessment is used, and a link for peer assessment submission instead of submitting a word document. There must be some rules imposed before submission, and it should be more private and confidential. Although it is not allowed to discuss the peer assessment, group members in higher education end up discussing it. Group members try to find out what other members have written about them, and then they try to penalize those members by giving low marks. The students must be allowed to have at least two self-selected team members. Participants also suggested consecutive peer assessment as there is always one person in a group who does not do anything in the beginning, but in the end, he does his work. This sort of behavior is unfair, and hence, consecutive peer assessment is the best option.

Participants also suggested that peer assessment can be taken as a short exam, for example, for half an hour, in front of any staff, not necessarily a module convener. Students should fill the peer assessment without cheating. The students should not know when the peer assessment is going to happen. If the students are taking more time filling up the peer assessment, then they are losing their chance to voice up. It may be better if the students do not know whether any group coursework contains peer assessment. It is better to use Moodle online submission instead of Word documents for peer assessment. In short, it should be online, promotive, at a random time, and within the fixed time. With these instructions, we may minimize the risk of people discussing, minimize friendship conflicts and fairness issues. At the end of the coursework, a clear breakdown of students' comments must be provided as a student has the right to know the scoring mechanism from peer assessment. Peer assessment should include detailed coursework allocation for each student which would ease the monitoring for the supervisors. Also, supervisors should divide the tasks among the group members in a way that they contain an equal distribution of the marks.

## 5. CONCLUSION

From the result of our study, we can conclude that an individual's perception of fairness, conflicts due to differences in opinion, and disruptive behavior negatively affecting the group dynamics are the main issues. Our suggestions to make group work a pleasant experience are by identifying and taking action against the freeloader, giving credit to the deserving students, educating students on how to give constructive feedback, and making the assessment process transparent to all. At the same time, making it easier for tutors and peers to give a formative assessment to students. The limitation of this research work is that this case study is only with a small specific population, whereas, the actual solutions may be more complicated, and may require additional specific support. Also, due to some reasons, the school did not have any Master's students at the time of the study, so the case study only involved undergraduates. In the future, similar studies could be conducted with postgraduate students as well to ensure the reliability of this research.

**Corresponding author:** Dr. Zhiyuan Chen
**Contact email:** zhiyuan.chen@nottingham.edu.my